*Attila Lajos Makai, Szabolcs Rámháp*

# The Changing Role of Entrepreneurial Universities in the Altering Innovation Policy: Opportunities Arising from the Paradigm Change in Light of the Experience of Széchenyi István University

*Summary*

The progress made by the entrepreneurial university, which is a newly emerging category in Hungarian higher education after its change of model, has not only deepened relations between universities and the industry and intensified the technology and knowledge transfer processes, but also increased the role of universities in shaping regional innovation policy. This transformation places co-operation between the actors of the regional innovation ecosystem and the relationships between the economic, governmental and academic systems into a new framework. The purpose of this paper is to describe the process of the change in the model through a specific example, and to outline the future possibilities of university involvement in the currently changing Hungarian innovation policy system.

Attila Lajos Makai PhD student, Széchenyi István University, Doctoral School for Regional and Economic Sciences (makai.attila.lajos@sze.hu); Dr Szabolcs Rámháp PhD, assistant professor, Széchenyi István University, Kautz Gyula Faculty of Economics (ramhap@ga.sze.hu).

---

## Introduction

Since the beginning of the 2000's development in the domestic innovation policy has been strongly influenced by Hungary's accession to the EU, as an external force, because the standards and structures enabling proper use of the available funding have not only influenced the methods of allocating resources, but also the system of policy institutions (Dőry, 2005). The massive regionalisation and decentralisation efforts behind the EU's normative system for the use of Structural Funds, which was also adopted by the Hungarian development policy and incorporated into the current regulations, can be said to be of particular importance. Decentralisation efforts had already appeared (during the period of adopting the Act XXI of 1996 on Regional Development), but they were implemented in the period of using the Pre-Accession Funds (PHARE, ISPA, SAPARD) and during the First National Development Plan (2004–2006). The planning and implementation of innovation policy has been strongly influenced not only by the National Strategic Reference Framework (NSRF, and later the Partnership Agreement) for the programmeming and use of EU funds, but also by the different strategies and organisational solutions of each political course. Besides the continuous expectation of decentralisation from the European Union, the implementation of the principle of decentralisation and centralisation has alternately characterised innovation policy in Hungary over the past 20 years. Overall, it can be stated that the individual EU budget periods did not provide a unified regulatory and organisational framework. In both of the periods between 2007 and 2013, and 2014 and 2020, there were significant and fundamental changes of direction, which set the surveyed policy on a radically different trajectory. The main features of innovation policy in each planning period were as follows:

1. 2007–2013

a) Emphasis on decentralisation in innovation policy

b) Influencing decentralised decision-making processes by sectoral and central interests

c) Involvement of regional organisations in policy implementation

d) The centralised nature of policy (and support) decision-making despite a centralised institutional system

e) Use of the definition of region in an administrative sense in the planning and implementation process; planning and implementation are not based on the concept of innovation ecosystems

f) Low intensity in technology transfer processes; the Hungarian Academy of Sciences has a significant role, while higher education institutions are relatively less important in shaping innovation policy

g) Shortage of venture capital funds, primacy of tender financing

2. Features, 2014–2020

a) Centralisation paradigm, emphasis on the national economic level

b) Influencing centralised decision-making processes by local interests

c) De-emphasising and effacing regional innovation organisations

d) The centralised nature of policy (and support) decision-making pathways

e) Launching the integration of the concept of regional innovation ecosystems into planning and implementation each programme

f) The strengthening of technology transfer processes and the business attitude of higher education institutions; the Hungarian Academy of Sciences continues to have a significant central role, whereas higher education institutions have a growing regional role in innovation policy

g) The emergence of state-financed or partially state-financed (e.g. JEREMIE-type) venture capital funds.

In spite of subsidies, Hungary has remained among the "moderately innovative" countries and the existing disadvantages have not been eliminated. Plans for a Partnership Agreement related to the budget to determine the EU funds for the period of 2021–2027 and the S3 strategy, which is also a priority for the innovation policy, are currently under way. However, the range and roles of the actors involved in the planning of innovation strategies and interventions described in each document differ significantly from the previous ones. Currently, universities, being the main innovation-organising institutions in each region, co-ordinate implementation of the outputs of regional actors to central governments. (A good example of this is the emergence and operation of the Territorial Innovation Platforms that underpin the new S3 strategy.) However, international examples and the literature show that only entrepreneurial universities, which are well embedded in innovation ecosystems, have a financial interest in co-operation and in the achievement of results, are able to perform third missions, and are suitable for the effective performance of these functions. Change in the university model and the above-mentioned planning process began at the same time in Hungary, and they are both expected to have a decisive impact on the innovation policy in the following years. The purpose of this study is to present the process of change in the university model through a practical example (of Széchenyi István University), and to describe the place and role of universities based on this new footing in Hungarian innovation policy.

## Changes in innovation policy after 2018: entrepreneurial universities, an entrepreneurial state and innovation ecosystems

Towards the end of the 2014–2020 programming period and of the preparatory work for the new budget period of 2021–2027, there has been a lively debate among the representatives of both the policy and the academic sub-system about the European and national issues of organising and supporting innovation. The shortcomings of the organisational structures seen in recent years, the effectiveness of the interventions implemented and the consequent not-so-encouraging innovation statistics confirm that there is a need for a radical change in the innovation policy and for the replacement of the usual dogmas and paradigms. The same process has been taking place in the European Commission, as at the EU level similar statistical data (Eurostat, OECD) prove that the goal set in 2013–2014 to reduce the European Union's lag in the field of R&D and innovation compared to the USA and Asian states, not only has not been achieved, but the gap that existed seven years ago has further expanded between Europe and its competitors. (A special problem was encountered as a result of the Brexit, entailing the secession of London's innovation and start-up ecosystem, which currently ranks 3rd globally and represents an ecosystem value higher than the aggregate value of the ecosystems of Berlin, Amsterdam and Copenhagen.)

Several solutions have been proposed at an EU level to resolve the situation. Within the frames of academic and political debates, the Commission has selected Mariana Mazzucato's proposal described as a "mission-oriented research and development strategy". Briefly, the proposal aims at a targeted co-operation between entrepreneurial states, entrepreneurial universities, and economic actors focusing on innovation and impact based on proactive local community networks through innovation development programmes, and implemented by the bottom-up initiatives defined by central (state and community) actors (Mazzucato, 2018a). The cited publication proposes interventions at Member State level, and in a subsequent strategic document, the author also makes recommendations for a European Community innovation policy for the next seven years (Mazzucato, 2019). However, mission-based planning requires a redefinition of the role of the public sector and use of the concept of economic value, which is done by Mazzucato in her comprehensive theoretical work (Mazzucato, 2018b). Strongly value-based theories, which are based on Keynes' theories and aim at redesigning the role of the public sector and increasing its significance (including Mazzucato's theory), have become increasingly popular since the great depression of 2008 (Skidelsky, 2018). Rethinking the role of the state (especially in innovation policy) is represented by the concept of an "entrepreneurial state", which appears to be a central element in the theory (Mazzucato, 2015). The author presents the operation of an entrepreneurial state through the basic and applied research funding activities of US government agencies, emphasising that this funding function is maintained in the commercial utilisation of the subject of the research by venture capital funds (partially) sustained and operated by the state. Venture capital is commonly seen exclusively as a fund set up by economic actors for long-term investment purposes, although it is not confirmed by either theoretical or empirical research. A good example is provided by Etzkowitz in his highly influential book titled *Triple Helix*, which describes the birth of the world's first venture capital fund at Harvard University, and then the adoption of the successful model at Stanford University, which highly contributed to the foundation of the subsequent success of Silicon Valley (Etzkowitz and Zhou, 2018). The claim that public equity funds are relatively new institutions and are mainly characteristic of Far Eastern innovation ecosystems is also incorrect. However, the reality is that the world's first state venture capital fund called In-Q-Tel was created by the CIA in the 1990's (Keller, 2016).

The mission-oriented research and development strategy has also had a significant influence in Hungary since 2018–2019. The new National Smart Specialisation Strategy for the period of 2021–2027 has already been prepared on the basis of this theory. The mission defined by the government (in line with the objectives of the Cohesion Policy) is to create a smart Europe by achieving the following general objectives:

– Strengthening RDI capacities and introducing advanced technologies;
– Digitisation for citizens, companies and governments;
– Strengthening competences and the entrepreneurship needed for S3;
– Enhancing the competitiveness and growth of SMEs.

Besides general objectives, the National Research, Development and Innovation Office (NRDIO), which is responsible for the preparation of the strategy, aims to identify regional strengths in Hungary and to utilise them as effectively as possible. The government plans to develop and then implement the new S3 strategy with the help of the Territorial Innovation Platforms (TIP) established in 2019, thus involving the regional level in the planning process, and creating an opportunity for decentralisation. The TIP structure does not have the previously used NUTS2-based regional planning frameworks but relies on innovation ecosystems as the basis of the territorial units of innovation policy (Nemzeti Kutatási, Fejlesztési és Innovációs Hivatal, 2020). The spatial centres of innovation ecosystems are universities which may become able to play the role of RIOs (Regional Innovation Organisers), a concept adopted by Etzkowitz, in a given innovation ecosystem with the help of academic entrepreneurship and entrepreneurial education functions. This, of course, requires a change in the model of strengthening the entrepreneurial attitude of universities, which is currently in progress parallel with the development of the S3 strategy and the planning of the 2021–2027 programming period. The planning and implementation of the process relies heavily on the experience of the first S3 strategy, the main elements of which were summarised by Imre Lengyel (Lengyel, 2018, p. 31).

– Subsidies granted to improve competitiveness and efficiency take precedence over the forms of subsidies based on social policy or equity, due to their economic growth-enhancing effects. Therefore, the interventions related to innovation policy should be included, to a far greater extent, in the operational programmes supporting regional development in the future.

– The process of spatial concentration is of importance for bottom-up and endogenous economic and business development. Based on this line of reasoning, it may be questionable whether the concept of region, in its administrative sense, is apt for planning interventions that strengthen emerging innovation networks in the future when using resources for the development of areas and regions.

– Without support to bottom-up initiatives that strengthen the background of innovative enterprises, developments may not be economically successful, whether in the public sector or in support of basic and applied research, since regionally embed-ded and agile early-stage enterprises are needed to explore their market potentials and the multi-directional utilisation potentials of the innovations.

– Improvement in the regional innovation ecosystems requires a local and embedded agency capable of adjusting national economic programmes to regional features and of representing regional demand at a sectoral level. All this requires either a decentralised administrative unit or a locally embedded work organisation capable of performing the above functions.

The mentioned experience also identifies the main functions of the entrepreneurial universities created by the paradigm change and operating as an integral part of the regional innovation ecosystem in course of the planning and implementation of innovation policy.

### Entrepreneurial universities

The de-emphasis of the traditional, dual higher education paradigm, well-known for Humboldt (Etzkowitz, 2019) was accompanied not only by an increase in the number of students in higher education, but also by an increase in the pace of technology transfer at universities. This phenomenon is well illustrated by the sharp increase in the rate of R&D expenditure in the US higher education system following the shift from the "science as resource" model to the "science as engine" model (Popp Berman, 2012). The establishment of service/entrepreneurial universities also means the penetration of economic logic into the academic/higher education system, resulting in "academic capitalism" (Münch, 2014), with its basic elements (with a special emphasis on inter-university rankings) facilitating the increasingly growing effect of economic logic in the continental, the British and the US-type higher education systems. The literature is not uniform in the definition, or concerning the functions and the features of an entrepreneurial university. This, of course, does not mean that there is no consensus in certain elements; the strong link between economic and academic systems and focus on innovation are not criticised by any trend or theory. As Etzkowitz definesit, "the entrepreneurial university, combining the 'third mission' of economic and social development, with teaching and research, is a growing contemporary phenomenon, in which academia takes leading role in an institutional base of an emerging mode of production based on continuing organisational and technological innovation" (Etzkowitz and Zhou, 2018, p. 58). In relation to this, Clark (1998, besides separating each university model) has previously described entrepreneurial universities' basic pillars as follows:

– Professional management;
– Improvement in university-related development peripherals;
– Diversification of funding;
– Academic background with a business incentive;
– Integrated entrepreneurial culture.

Within the framework of the Triple Helix model, the co-operation between universities, economic and scientific systems creates the preconditions of effective innova-tion, and more specifically, the university as the location and integrator of innovation provides the complex factor (Leydesdorff and Etzkowitz, 2001), suitable for creating diffuse and interactive innovation processes in the platform of open innovation (Chesbrough, 2003). A further development of these ideas is the Quadruple Helix model developed by Carayannis and Campbell (2009), where the fourth "spiral" is provided by the public and the local media. A few years later, the same authors identified the environmental and social challenges associated with global warming as two of the most important drivers of innovation. These, according to them, are so important that they are incorporated into the innovation model as a "fifth spiral", creating the Quintuple Helix model (Carayannis et al., 2012). Regarding the latter theory, the arbitrary emphasis of the effects of global warming, which seems to be a kind of value choice rather than the objective description of the phenomenon, may arise as a criticism. As examples, it is enough to mention the challenges of globalisation, the growth of various inequalities, or human/social adaptation to the acceleration of technological development. Each of these may become a "spiral" in a given space and time, thus allowing the model to be arbitrarily expanded to a Sextuple Helix, and then to an Octuple Helix direction, which rather makes scientific observation difficult than giving it a tool. These circumstances are factors that determine the direction and nature of innovation rather than its organisational and policy framework. Therefore, to maintain a clear and practical theoretical framework, on the one hand, the Triple Helix model can be effectively viewed using the tools of Luhmann's system theory (Luhmann, 2012), and each "spiral" can be interpreted as Luhmann's social subsystem, while on the other hand, their interaction may be interpreted as described by Münch (2011).

Of course, a university as an actor existing in space has a strong influence on the economic processes of its narrower and wider environment. Chatterton and Goddard (2000) present the relationship between universities and the regional economy, and they emphasise interdependence in their model: a strong and innovative regional economy supports the position of universities, and a strong university can be a catalyst for development and innovation in a region. The Chatterton-Goddard model, unlike the Triple Helix models, is built on regionalism and decentralisation. The model links the role of universities to the economic stimulus of regional policy. In this concept, knowledge networks are essential elements of regional economic development, and universities are not only sources of innovation, but also represent the nodes in knowledge networks (Rámháp, 2018, p. 375). In relation to this, the concept of "learning region" (Hassink, 2001) deserves mentioning, as it provides a framework for the organisational, cultural and institutional analysis of knowledge-based regional innovation development. This paradigm also describes regional higher education institutions as key actors in regional development and innovation (Rutten et al., 2003). The models emphasising the regional role and impact of universities and decentralisation, and the Triple Helix model are in fact compatible with each other. One model emphasises the regional missions of higher education institutions, and the other outlines the optimal operational structure in which the actors of each subsystem can co-operate optimally to develop regional innovation ecosystems. The factors and effects of regionalism and regional development policy can be incorporated into the analytical framework of the Triple Helix model and can also be analysed (Ranga and Etzkowitz, 2015). All this is supported by Hungarian research, which reveals that the appropriate allocation of economic, cultural and social capital and co-operation between regional and state actors according to the Triple (Quadruple) Helix create an opportunity for the development of knowledge regions (Rechnitzer, 2016, p. 245).

Wissema (2009) derives the functions of a third-generation university through the model of Cambridge University, considered as a benchmark. It focuses on the process unfolding since the 1970's and in the course of which, thanks to the institution, a high-tech industry has developed around Cambridge (Cambridge phenomenon). Wissema has identified seven main features: (1) the formulation of the "third objective": the utilisation and dissemination of knowledge will be central; (2) operation in an international, competitive market will be a priority; (3) universities will become open to collaboration with industrial and business partners; (4) trans- and multidisciplinary research will come into prominence as the nature of research changes; (5) both mass and elite education will be undertaken; (6) institutions will become multicultural organisations; and (7) no direct state financing and no state interference will be made (Wissema, 2009).

The development of innovation ecosystems in Central and Eastern Europe has followed the model of the key countries of Europe with a significant time lag, but, of course, the fact that the regions in this area are considered peripheral in terms of innovation also has a significant impact. This is also supported by the regional innovation indicators calculated on the basis of the Regional Innovation Scoreboard (RIS) framework, which show a significant lag behind the regions leading these areas (e.g. the western and southern regions of Germany, the Netherlands, the southern regions of the UK, and Ireland) in all cases since 2007 (Gál and Páger, 2018). Another feature is the continuous decline in the number of students in higher education (with the exception of the Czech Republic and Croatia), and the key role of state interference in the formation of innovation processes. Another characteristic of the region's innovation ecosystems is that the regional innovation organisers (RIOs), which are needed for the effective operation of the Triple Helix model and which are able to organise, co-ordinate and operate the region's innovation, knowledge and consensus spaces, cannot be identified in them or only to a limited extent (Ranga and Etzkowitz, 2015, p. 124). This function has been increasingly taken over from the various actors of regional policy by individual higher education institutions that will become suitable for this role after transformation into entrepreneurial universities. Based on the research of Nieth and Bennewoth (2019), each local university contributes to the development of each peripheral region, inter alia, by mediating between demand for and the supply of innovation, providing supportive infrastructure, building the necessary regional partnerships, and developing the necessary skills.

304

## Széchenyi István University becoming an entrepreneurial university

*History*

Since 2010 integration has taken place in Hungarian higher education due to economic necessity. The Structural Reform Plan withdrew significant funds from the system: HUF 12 billion, HUF 38 billion and HUF 28 billion were withdrawn from higher education in 2012, 2013 and 2014, respectively, representing a reduction of 6 per cent in 2012, and 19 per cent in each of 2013 and 2014, compared to 2010. These cuts were offset by the Human Resources Development Operational Programme (HRDOP), which opened a wide range of tender opportunities for institutions. The purpose of the HRDOP is to increase the ratio of higher education graduates by improving the quality and accessibility of higher education, to improve the labour market relevance of training, to ensure the supply of researchers, and to increase contribution to Hungary's economic development objectives through the Structural Funds (ESF and ERDF). Integration and economic necessity have had two effects:

– the regional weight and role of integrated institutions has increased, and

– the withdrawal of funds and the use of tender funds earmarked as a replacement have developed and subsequently strengthened business and entrepreneurial approach by universities.

The draft of the government's new higher education strategy entitled "Graduation Change in Higher Education. Guidelines for Higher Education Development with Focus on Performance" was published in October 2014 under the auspices of the Ministry of Human Capacities. The strategy adopted by the government in December of the same year specified guidelines for the development of the Hungarian higher education system for the period between 2015 and 2030.

The objectives of the document relevant to innovation policy are as follows:

1. To develop a higher education system focussed on performance and to improve the quality of higher education;

2. To strengthen research elements within higher education and to make research careers more attractive;

3. Closer links between local/regional enterprises and higher education institutions.

As a first step of implementing the strategy, the amendment to Act CCIV of 2011 on National Higher Education and its enacting order adopted a new training structure including community-based higher education centres, chancelleries and a consistory system.

*Paradigm change at Széchenyi István University*

Széchenyi István University is the largest higher education institution in the North Transdanubia region, Hungary's most developed agricultural and food industrial

305

area in the country, where industry nevertheless plays the leading role and has been growing rapidly since the second half of the 1990's. Due to its geographical location, the northern part of the region – mainly Győr-Moson-Sopron County, and to a lesser extent Vas County – is actively involved in the circulation of Europe and the world market, with its comparative advantage still provided by industry (mainly construction and vehicle industry). GDP per capita in the region exceeds the national average (Tamándl et al., 2014).

The predecessor of the university, called the Technical College of Transport and Telecommunications, was founded in 1968, with the aim to train the technical intelligentsia needed for the development and maintenance of transport and telecommunications infrastructure in the region. Its name was changed to Széchenyi István College in 1986. Since the 1990's in addition to engineering, health sciences, economics, law and arts have also been added to the subjects taught here. The expansion allowed the institution to gain a university rank in 2002. The transformation to an entrepreneurial university was greatly boosted by the objectives of the medium- and long-term development concepts of the city of Győr, home to the university. Based on these objectives, it was set as an aim that the city should transform from a vehicle manufacturing location to a development centre which contributes to the evolution of a competitive regional economy by developing and diversifying knowledge industry (Fekete and Rechnitzer, 2019).

The most important tool in achieving the objective was capacity building at Széchenyi István University, carried out in several stages with the financial support of the European Union. The core elements of the programmes include the ability to innovate, the development of human and infrastructural potential needed for the efficient implementation of R&D processes, which ensure the efficient, two-way flow of knowledge and information between the economic and university sub-systems.

A next step in transformation into entrepreneurial university was launching the university's Centre for Co-operation between the Higher Education and the Industry and the infrastructural investments began in July 2017. Within the framework of the project, only three buildings (Management Campus, Logistics Packaging Laboratory, Brake pad Testing Building) were built or their infrastructure improved. The Management Campus, inaugurated in 2018 and established with funding from the Economic Development and Innovation Operational Programme (EDIOP), is home to the university's entrepreneurial and technology transfer functions, and operates as an individual competence centre within the university. In the project, the achievement of two lasting results are planned: 1. The ultimate aim of the project is to elaborate a portfolio of services for SMEs by the Centre for Co-operation between the Higher Education and the Industry (FIEK), adapted to the local economic environment, and representing a complex product-package consisting of training, product and organisational development services and operation according to industrial order. 2. The Management Campus, being a research and training institution, works out original and forward-looking scientific products, and publishes them in international publications (Eisingerné Balassa and Rámháp, 2019).

The aim of this centre is to conduct research in organisation and product development and problems in large companies and SMEs and to answer them on a scientific basis. Within this framework, in addition to scientific research and analysis, a portfolio of services closely adapted to the local economic environment is created, thus it is suitable for promoting close and long-term co-operation between the university and economic actors. The Management Campus basically contributes to the strengthening of entrepreneurial university functions by implementing the following activities (Széchenyi István University, 2018):

– Student innovation projects;
– Innovation services;
– Innovation promotion;
– Supporting early-stage enterprises (start-up and spin-off companies).

The most recent important step in the transformation to an entrepreneurial university is the process known as a paradigm change in higher education, aimed at creating an efficient and modern higher education. During the process, the fundatorial and controlling rights of paradigm-changing institutions are handed over by the state to trust foundations. All this is an important step towards developing the ability to innovate and opening up to the business world. As a result of the process, on 1 August 2020 the following universities changed: University of Veterinary Medicine Budapest, Moholy-Nagy University of Art and Design Foundation, University of Miskolc, John von Neumann University, Széchenyi István University and University of Sopron; and the University of Theatre and Film Arts may join them as the seventh paradigm-changing university. The controlling rights of Széchenyi István University were transferred from the state to the Széchenyi István University Foundation. As the open letter of the Board of Trustees of the Széchenyi István University Foundation states: "In recent years, Széchenyi István University has become one of the leading higher education institutions in the region and in Hungary, has significantly increased the number of its students and its revenues, and has improved its infrastructure, with its stability ensured by its outstanding liquidity. All this provides a solid foundation for the university to become a winner in the renewal opportunity now opening up". As a result of the process, a more efficient and flexible operation, greater independence, new developments, and the further strengthening of the university's corporate partnerships, internationalisation and service capacity are expected (Széchenyi István University, 2020b).

## Venture capital in the university-centred innovation ecosystem

Venture funds are currently characterised by capital deriving partly or fully from public sources, so the number and ratio of private equity funds are very slight compared to public funds, and the participation of private investors is low (Kormányzati Informatikai Fejlesztési Ügynökség, 2018). Since the beginning of the 2000's approximately HUF 200 billion has been made available for public funds for venture capital investments, and some 500/600 Hungarian enterprises have received venture capital from this overall amount. Development of the Hungarian start-up ecosystem began to develop mainly on launching the JEREMIE funds, financed from EU funds in the period between 2007 and 2013. The amount of funds available for them exceeded HUF 85 billion (Karsai, 2013). In the EU's programming period 2007-2013, Hungary, for the first time, had the opportunity to use a larger amount of the EU funds in the form of venture capital. However, according to Karsai (2017), the relevant EU regulation significantly limited the spatial and temporal use of these funds, and these limitations were difficult to reconcile with the nature and time-demand of venture capital investments. Moreover, an extensive set of rules to avoid illegal state aid also narrowed the room for manoeuvre for fund managers who provide venture capital to enterprises. All these regulations were equally unusual in the domestic market for public administration, investors, and enterprises looking for venture capital. In the period 2014–2020, venture capital funds with state participation targeted the placement of capital worth a total of HUF 250–260 billion (Karsai, 2017). Higher education institutions are also important partners to this.

Hungarian venture capital funds and the start-up ecosystem are centred in Budapest, as evidenced by the facts that the headquarters and the seats of all the fund are in Budapest, and that a significant part of the start-ups that have received investment are also based in Budapest (Jáki et al., 2019). Strong concentration in the capital city is somewhat mitigated by the EU's special rules that allow the use of certain funding sources only in convergence regions, thus in case of some funds, only non-capital start-ups can expect investment, given that these funds use this type of EU funding. *Inter alia*, the allocation requirement resulted in some capital funds' appearance in the country's major cities outside Budapest. Based on the experience of the past two years, it can be said that a significant part of the applications for investments can be found in the major cities with campuses (e.g. Miskolc, Debrecen, Győr) (Jáki et al., 2019). Therefore, it is not a coincidence that some capital funds, especially the Hiventures state capital fund, which has the most significant capital and apparatus in Hungary, have devoted increasing serious resources to ensure their presence in university campuses. Co-operation in the field of innovation is also supported by the government, the most recent example being the establishment of Territorial Innovation Platforms in Győr in November 2019 (Széchenyi István University, 2019).

*Venture capital funds at Széchenyi István University*

Embracing student innovations began about 10 years ago at Széchenyi István University. The Knowledge Management Centre, which also carries out technology transfer functions, was established at the institution in 2010. It was around this time that the Spin-off Club, organising meetings with successful entrepreneurs, and the Business from Idea initiative, which has been enjoying unbroken popularity among the students of engineering ever since, were launched to encourage entrepreneurship. Students are required to build a business model based on their own ideas by the end of a given term. The course focuses on practice, as the university lecturer is assisted by two successful local entrepreneurs, who give practical examples and share their experience. The SZE-Duó competition is also rooted here; within its frames one lecturer and 2 or 3 students were given the opportunity (under a contract for services) to create a prototype: there were 3 supported projects in every term for 4 years, financed through tenders and from the university's funds, and several successful start-ups were set up. The aim of the Creative Summer Universities initiative, organised in 2014 and 2015, was to develop business ideas in groups of 2/4 people, whose work was assisted by designers, marketing professionals and business development mentors. As a result, the university entered the national start-up cycle. Széchenyi István University and Quantum Leap incubator formed an alliance in 2017 within the framework of the GINOP-2.1.5-15. project titled "Innovation Ecosystem (start-up and spin-off)". As a result, the first university incubator was established, which supported these groups by training and investment from tender sources. The aim of the project was to support innovative start-ups from the birth of an idea through validation and prototype development to market entry, developing a sustainable business model and achieving suitability for investment. Consequently, in the academic year 2017–2018 incubation workshops were launched, and the university came into contact with Hiventures capital fund. In 2019, under the trilateral agreement between Széchenyi István University, Hiventures Zrt. and Enterprise Hungary Nonprofit Kft., the Startup Campus Győr incubation programme was established, and launched a more complex training and investment programme. The Input programme launched in the meantime assisted start-ups in Győr by training courses and by a mentoring network, including enthusiastic volunteers who started to build a community and organise events called Start Up Wednesdays.

At Széchenyi István University, as in most technical universities taken in the classical sense of the word, communication with the start-up ecosystem basically takes place on the Management Campus, which co-ordinates technology transfer processes and was estalished from FIEK resources. The university and the Management Campus co-operate with the mentioned Hiventures capital fund in start-up incubation, with Quantum Leap that also carries out incubation and preseed investment functions (Quantum Leap, 2020b), and also with Startup Campus incubators. The jointly implemented (non-capital investment) programmes are the following (Széchenyi István University, 2020a):

– Start-up consultation hours;
– Start-up incubation workshop;
– Mentoring;
– Co-working office on the Management Campus;
– SZE-Duó competition;
– Spin-off Club.

The incubation programme implemented by Quantum Leap using EU funds (Quantum Leap, 2020b) is a particularly forward-looking initiative in Hungary, and in addition to acting as a partner, Széchenyi István University also proactively participates in the investment (and decision-making) process through the Management Campus. This is how it provides the human resources involved in the provision of technology transfer processes with the competencies and experience that will be used in the future in order to establish and operate an independent university venture capital fund. The establishment of a possible university capital fund may be facilitated by the paradigm change planned at Széchenyi István University and launched in August 2020, with control transferred from the state to a foundation entailing release of the institution from the public finance sub-system, which will promote the fulfilment of the entrepreneurial functions of the university by a positive change in the relevant standards.

## Conclusions

The emergence of new stakeholders (entrepreneurial universities and venture capital funds) in the planning and implementation of innovation policy and the new goals to be achieved and the interventions helping to achieve them have induced the following changes:

1. Mission-driven central innovation policy planning (entrepreneurial state);

2. Strengthening the role of regional innovation ecosystems (networks of ecosystems);

3. Entrepreneurial turn and the paradigm change in case of universities forming the centre of the local innovation space (entrepreneurial university).

In relation to this, entrepreneurial universities, acting as regional innovation-organising ecosystem nodes after the paradigm change, can contribute to the following functions:

– Contribute to strengthening the portfolio of state and quasi-state venture capital funds related to a given region and to adapting to the regional industry profile promoting this way the "smart money" nature of each capital fund.

– Relying on the professional capacities developed at entrepreneurial universities and participate in organising regional innovation ecosystem, universities can effectively take over and fulfil the functions of former regional innovation agencies.

– The universities' international relations and networks that have traditionally existed or developed as a result of previous internationalisation projects may contribute to developing the global visibility of regional innovation ecosystems.

In Hungary the process of the paradigm change at universities began in 2019, and were completed in 2020. By the first months of 2021 the strategies and planning documents intended to determine the support system in the next seven years will have been finalised. The effects and efficiency of changes in the renewed innovation policy cannot yet be measured or analysed. However, the directions and depth of changes are clear, and based on this it can be stated that in the next seven years the funding and application system supporting innovation, and the organisational structure carrying out its implementation and co-ordination will be fundamentally different, but the entrepreneurial universities established after the paradigm change can make a meaningful contribution.

311

313